\documentstyle[aps,latexsym,graphicx]{revtex}
\newcommand{\eqref}[1]{(\ref{#1})}

\newcommand{\pa}{)\ }

\newcommand{\be}{\begin{equation}}
\newcommand{\ee}{\end{equation}}

\begin{document}

\title{Ising-like dynamical signatures \\
and the end-point of the QCD transition line}
\author{
Sz. Bors\'anyi$^{ }$\footnote{e-mail: mazsx@cleopatra.elte.hu}, A.
Patk{\'o}s$^{ }$\footnote{e-mail: patkos@ludens.elte.hu}, D. Sexty$^{
}$\footnote{e-mail: sexty@cleopatra.elte.hu}, Zs. Sz{\'e}p$^{
}$\footnote{e-mail: szepzs@cleopatra.elte.hu}\\
 Department of Atomic Physics, E{\"o}tv{\"o}s University, H-1117 Budapest,
Hungary}
\maketitle
\begin{abstract}
An increase in the size of coherent domains in the one component $\Phi^4$
field theory under the influence of a uniformly changing external magnetic
field near the critical end-point $T_{\Phi}=T_c, h_{\Phi}=0$ was proposed 
recently as an estimate also for the variation of the chiral
correlation length
of QCD near its respective hypothetical end point in the $T_{QCD}-\mu_{QCD}$
plane.  The present detailed numerical investigation of the effective model
suggests that passing by the critical QCD end point with realistic rate of
temperature change will trigger large amplitude oscillations in the temporal
variation of the chiral correlation length.  A simple mechanism for
producing this phenomenon is suggested.
\end{abstract}
\section{Introduction}

The quality of our knowledge of the phase structure of QCD at high
temperature and finite baryon density is an important benchmark for our
understanding of strong interactions.

A critical end-point of the first order phase transition line in the
$T-\mu$-projection of the QCD phase diagram was conjectured 
\cite{stephanov98} to follow from the compatibility of the following 
observations:

i) Lattice simulations \cite{karsch01} indicate that the phase
transformation at zero chemical potential with realistic quark masses is a
crossover, characterized by an analytic variation of the thermodynamical
potential with the temperature;

ii) At zero temperature there is a first order phase transition from the
hadron phase to more exotic phases as a function of $\mu$
\cite{barducci89} which continues as a first order line into the 
$(T,\mu )$-plane.

For the strong matter the significance of this end-point would be similar to
that of the Curie point for ferromagnets.

Recently, some important progress was realized in the search for the
 $(\mu_E,T_E)$ location of the end-point with non-perturbative lattice 
studies \cite{katz01}. There is a chance in current heavy ion collision
experiments that it can be observed experimentally, since with increasing 
collision energy/nucleon the central rapidity particle spectra explore 
regions of the phase diagram corresponding to decreasing chemical potential. 
An obvious class of the signatures would reflect the increasing size of the 
coherent
fluctuations in the order parameter $\sigma$-field in the neighborhood of 
the end-point \cite{stephanov99}. In this region mostly the $\sigma$-field
will be excited, since its mass is the lightest near the critical end-point
in view of the amount of explicit chiral symmetry breaking which keeps the
pions massive. This coherence should be reflected by the statistics of the
main decay products of the $\sigma$-field, the pions.

Starting from equilibrium at some $T_0>T_c$, and passing with finite velocity
near the end point, the system unavoidably slows out of equilibrium.  In
contrast to the equilibrium characterization of the second order transition,
finite maximal correlation length is reached with a certain amount of
supercooling. A substantial increase in the correlation length will be a
reliable signal for the existence of the critical end-point.

Quantitative predictions for this phenomenon should rely on non-equilibrium
field theory. For the moment it is hopeless to simulate directly the far
from equilibrium behavior of QCD matter. One recognizes however, that the
mass of the $\sigma$ in this region is separated by a gap from the other
mesonic excitations. Therefore, one arrives at the conclusion that an
effective dynamical theory of the longest wavelength excitations in this
region is in the same universality class as the Ising model. At present no
quantitative matching is known between the original theory and its
one-component $\Phi^4$ effective version. Still, universal features of
the class A of critical dynamics (in the classification of Hohenberg and
Halperin \cite{hohenberg77}) are expected to occur. \footnote{A more complete 
theory reflecting the direct influence of the temporal variation of a
conserved baryon number density on the $\sigma$ field falls into class C
\cite{berdnikov00}, and will be subject of future study.}

Recently Berdnikov and Rajagopal (BR) \cite{berdnikov00} proposed an
intuitive mapping. They approximately identify the magnetic field ($h$) of
the Ising model with the temperature of the QCD (the Ising reduced
temperature ($r$) axis is nearly parallel to the chemical potential axis of
QCD). They have checked that the results are not sensitive to a moderate
tilt in this mapping. Next, they have proposed a dynamical equation which
describes the evolution of the inverse correlation length (the mass of the
$\sigma$ meson) when the system passes through the critical end point of the
Ising model with finite velocity under different angles. This equation was
shown to depend on a single non-universal parameter, proportional to the
rate of change of the Ising magnetic field.

Finally, a semi-quantitative correspondence was proposed between the
relevant QCD temperature range ($T_0=180$~MeV, $T_{freezeout}=120$~MeV) and
the dimensionless strength of the Ising magnetic field
$(h_0=-0.2, h_{final}=0.1)$. The non-universal constant was varied in a wide 
range, since it relates essentially the rate of cooling of the QCD-matter to 
the speed of the variation of the external field $h$. The main result of 
Ref.\cite{berdnikov00} was a prediction for the variation  of the
order parameter (e.g. $\sigma$) correlation length with the variation of $h$
in the neighborhood of the critical end-point. 

The variation of the correlation length during the passage through the
critical point was compared to its value taken at the non-critical starting
value, which corresponds to the equilibrium energy density of the QCD plasma
produced in the collision ($T=T_0$).  It is the relative increase in the
correlation length at the freeze-out of the system ($T=T_{freezeout}$), where
most of the pions are coming from, which is the most important issue. A
factor of 2-3 increase was signalled, which is claimed to be observable
under the accuracy of the present heavy ion experiments.

The aim of our present investigation is to check the accuracy of the first
order relaxational dynamics assumed for the inverse correlation length. For
this we have integrated exactly the equations of motion of the one-component
classical three-dimensional scalar field, $\Phi_d({\bf x},t)$ on lattices of
size $N=16-128$:
\be
\ddot\Phi_d ({\bf x},t)=\left (\Delta -m^2\right )\Phi_d ({\bf x},t)
-{\lambda\over 6}\Phi_d^3({\bf x},t)-h_d(t),
\label{fulleom}
\ee
and measured simultaneously the variation of the order parameter and of the
correlation length.

Though the hadronic system freezes out at $T_{freezeout}\sim 120$~MeV which
corresponds to $h=0.1$ in the effective system, we have followed the
variation of the correlation length and of the order parameter to higher
values of $h$. This enabled us to recognize the relevance of a second order
dynamics in the effective equation of motion of these characteristics.

It turns out that the order parameter (OP) obeys an equation which is
slightly more complicated than the one proposed by Halperin and Hohenberg
for this class. It is formally analogous to the differential equation of a
damped oscillator. In order to achieve good quantitative description of the
OP-trajectory obtained from the numerical simulations, one has to take into
account the effect of slowing out from equilibrium while the system passes
by the critical point.

Our paper is organized the following way.  In Section II the method of the
numerical solution of (\ref{fulleom}) is shortly outlined. The methods of
analyzing the time evolution of the system are presented in more detail with
special emphasis on the determination of the spatial correlation length. In
Section III the trajectories of the most important quantities characterizing
the state of the system are mapped out in the $(r,h)$- and the $({\textrm
OP},h)$-planes when passing at different distances by the critical Ising end
point.  In Section IV we present the results for the non-equilibrium
$h$-evolution of the correlation length for several values of
$a^{-1}\equiv dh/dt$ and compare them with the estimates of BR. Finally, in
Section V a quantitative phenomenological interpretation of the measured
order parameter trajectory is offered.  Our dynamical description is
compared to the equation of motion proposed for this class by Halperin and
Hohenberg. Based on the proposed effective relaxational dynamics we suggest
also a simple way to account for the variation of the correlation length. 
Conclusions of our investigation are summarized in Section VI.

\section{Methods of solving and analyzing the exact classical equations}

For setting up and solving numerically the $\Phi^4$ theory we used
techniques similar to those applied in our previous paper \cite{borsanyi00}. 
Here we shortly outline the procedure.

As a first step one has to rewrite Eq.(\ref{fulleom}) using dimensionless
quantities (not having index $d$), defined as follows:
\begin{eqnarray}
&
t=t_d/a_x,\qquad x=x_d/a_x,\nonumber\\
&
\Phi =\sqrt{\lambda\over 6}a_x\Phi_d,\qquad
h=\sqrt{\lambda\over 6}{a_x^3}h_d,
\label{rescale}
\end{eqnarray}
where $a_x$ stands for the lattice spacing, the powers of which we used to
scale dimensionfull quantities. The dimensionless mass parameter of the
theory $|m|=a_x|m_d|$ was set to unity. The field equation in dimensionless
variables is of the following form:
\begin{eqnarray}
\Phi_{\bf n}(t+a_t)+\Phi_{\bf n}(t-a_t)-2\Phi_{\bf n}(t)&-&
{a_t^2/a_x^2}\sum_i(\Phi_{\bf n+\hat i}(t)+\Phi_{\bf n-
\hat i}(t)-2\Phi_{\bf n}(t))
+
a_t^2(-\Phi_{\bf n}+\Phi_{\bf n}^3-h(t))=0.
\label{fieldeom}
\end{eqnarray}

There were a number of simple quantities routinely monitored in each run.
The first was twice the average kinetic energy per site (called kinetic
temperature also for non-equilibrium field configurations):
\be
T_{kin}(t)={1\over L^3}\sum_{\bf n}\dot\Phi_{\bf n}^2(t),
\ee
the second, the trajectory of the homogeneous (OP) mode: 
\be
\bar\Phi(t)={1\over L^3}\sum_{\bf n}\Phi_{\bf n}(t)\equiv
\overline{\Phi(x,t)}^V.
\ee
Also its fluctuation 
\be
\delta\Phi^2(t)=\overline{\Phi^2(x,t)}^V-
\left(\overline{\Phi(x,t)}^V\right)^2
\label{flucteq}
\ee
was used.

The thermalization algorithm, which led to the initial state, consisted of
two steps. First we have set the initial energy density by continuously
comparing the desired and the measured kinetic temperatures. Depending on
the deviation from the targeted temperature an artificial friction or
anti-friction term was introduced into Eq.(\ref{fieldeom}). After having
reached the required kinetic energy density, in a second step the original
form of the discretized nonlinear field equation was iterated until
thermalization was complete. Both steps were repeated until the final
kinetic temperature was just what we desired.

The critical temperature $T_{\Phi,c}$ was determined at $h=0$ by locating
on the $T_{\Phi}$ axis the maximum of $\delta\Phi^2$, or of the specific
heat, and by finding the point separating zero and non-zero expectation
value regions of OP, $\bar\Phi$. The reduced temperature
$r\equiv (T_{\Phi}-T_{\Phi,c})/T_{\Phi,c}$ was measured relative to
$T_{\Phi,c}=1.5$ for $L=32$ and $T_{\Phi,c}=1.57$ for $L=64$.

\subsection{Determination of the correlation length}

The main goal of this Paper is to give a quantitative interpretation to the
dynamical behavior of the correlation length as the system passes by the
critical end point. Therefore we need an accurate measurement method for
this quantity, which is reliable in a dynamical system too.

\newcounter{directalgnum}
\begin{enumerate}

\item\label{directcorr}
The most straightforward way is to use the definition of the correlation
function:
\be
C(\Delta ,t)=\langle\overline{\Phi (x,y,z+\Delta,t)\Phi (z,y,z,t)}\rangle -
\langle\bar\Phi (t)\rangle^2.
\ee
Here the overline refers to the spatial average of some quantity at fixed
$t$ in a single sample, while the brackets stand for the average over the
initial conditions. After checking that $C(\Delta ,t)$ truly behaves as
\hbox{$\sim\cosh [(\Delta -L/2)/\xi_1 ]$}, one can extract the correlation 
length $\xi_1$. We refer to this characteristic length as the ``direct'' 
length of correlation below.

\item\label{grainsize}
Average linear size of the regions of same sign deviations from the space 
average $\bar\Phi$ can be taken to estimate the characteristic size of 
coherent fluctuations. Consider $\Phi ({\bf x},t)-
\bar\Phi (t)$. At a given time a histogram was constructed by scanning through
the lattice for the number of occurrences of site sequences with the same sign
deviation from $\bar\Phi (t)$, and of a given length 
$\Delta$, parallel to the lattice axes. The histogram showed perfect 
exponential dependence on $\Delta$. Its characteristic length defines
the ``grain size'' $\xi_2$. Repeating this measurement for every 
configuration during the time evolution one obtains the function $\xi_2(t)$.

\setcounter{directalgnum}{\theenumi}
\end{enumerate}

The correlation lengths defined by the above algorithms are different, of
course. One expects, however, that both definitions capture the same feature
of a field configuration and there exist simple functional relationships
between them. In order to find the relation of $\xi_i$ to 
the standard correlation length (or its inverse, the mass), which is usually 
measured with method \ref{resummedmethod} (see below), we studied the 
equilibrium systems at different values of the reduced temperature $r\in
(0,0.27)$. Using the algorithms described above one finds the
functions $\xi_i(r)$, as well the standardly used $m_{eff}(r)$. The 
elimination of $r$ leads to the relations $m_{eff}=g_i(\xi_i)$, well
approximated by second order polynomials. In principle, one should establish
this relation for each value of $h$ separately, but in practice the
relation found for $h=0$ gave good agreement between the results of
the different methods, also away from equilibrium.

\subsection{Spectral determination of the mass}

An algorithm for the reconstruction of the effective potential from the real
time evolution of a scalar field was presented in \cite{borsanyi00}. An
effective equation of motion was fitted to the temporal variation of the
order parameter. Its ``force'' term was interpreted as the derivative of the
effective potential with respect to the field. Then it was easy to identify
the effective squared mass.

This time we improved further this algorithm. All spatial Fourier modes of
the system are used for the reconstruction of the dispersion relations.
Three stages of the implementation were worked out.

\begin{enumerate}
\setcounter{enumi}{\thedirectalgnum}
\item \label{loceqmethod}
We assumed the validity of the following approximate equation for short time 
intervals:
\be
Z\ddot\Phi - \triangle \Phi + m_{eff}^2\Phi + \epsilon\triangle^2\Phi = 0,
\label{loceq}
\ee
where $\triangle$ is understood as lattice discretization to the Laplacian.
Comparing this to the real nonlinear evolution we could fit
$Z,m_{eff}^2,\epsilon$ for the short intervals in question in the following
way. A spatial FFT algorithm was applied to the field configurations 
generated by Eq.(\ref{fieldeom}). The temporal trajectory of each $\hat{\bf k}$
mode was fitted to the equation dictated by the Fourier transform of 
Eq.(\ref{loceq}). The coefficients of the polynomial of ${\hat{k}}^2$ 
determine also the time dependent effective mass. (Here and in the following
$\hat{k}_i=2(\sin k_i/2)$ stands for the dimensionless lattice
momentum.) A lower bound for the length of time interval
in which Eq. (\ref{loceq}) is fitted comes from the consistency
criterion that the time interval we averaged over must be much greater
than the resulting $1/m_{eff}$ time scale. For out-of-equilibrium
field configurations there is also an upper bound for the time
interval which is set by the variation rate of the parameters $r,h$.

\item \label{ksmethod} 
For the two-point function related to the OP-susceptibility \cite{rajagopal93}
the relation
\begin{equation}
\frac{T}{\left\langle{|\Phi_{\bf k}|^2}\right\rangle}
=Z^{-1}\left(m_{eff}^2+{\hat k}^2+\varepsilon{\hat k}^4+
{\mathcal O}({\hat k}^6)\right)
\label{twopointeffmass}
\end{equation}
holds in equilibrium \cite{tang98}.
Replacing the ensemble average in the denominator of the left hand
side by averaging over the $\bf k$-modes characterized by equal
$\hat{k}^2$ and using 
$T_{kin,\hat{k}^2}=\overline{|\dot\Phi_{\bf k}|^2}^{\Omega_{\bf k}}$ 
for $T$ we found that our system obeys Eq.(\ref{twopointeffmass}) for 
${\hat {k}}^2>1$ without any time averaging. With these replacements we
could fit the value of $m_{eff}^2(h)$ continuously both near and far
from equilibrium. The wave function renormalization constant was found to be
equal to 1 up to 0.5 percent in all cases. The coefficient $\varepsilon$ 
of the fourth derivative correction term was checked to be  negligible. 
\item \label{resummedmethod}
Since most of modes with ${\hat {k}}^2>1$ nicely follow the above behavior,
one might improve the statistical confidence of the above algorithm by
summing over the contribution of all modes (including $\hat{k}^2<1$),
which yields the approximate equality
\begin{equation}
\frac{\delta\Phi^2}{T}=\frac1{L^3}\sum_{\bf k\ne 0}\frac{1}{m_{eff}^2+
{\hat k}^2},
\label{meffeq}
\end{equation}
with $\delta\Phi^2$ denoting the field fluctuation as defined in
Eq.(\ref{flucteq}). Its measurement does not require the application of the
time consuming FFT. The value of the function on the right hand side can be
tabulated and Eq.(\ref{meffeq}) is quickly solved with help of a look-up
table plus interpolation for every measured value of $\delta\Phi^2$.
\end{enumerate}

In addition, as a quick reference the so-called Hartree squared mass estimate
$m^2_{Hartree}=-1+3({\bar\Phi}^2+\delta\Phi^2)$ was also used. It proved 
useful in interpreting qualitatively effects in the motion of modes with 
different $\bf\hat k$ and apparently related to the presence of an 
$h$-dependent common  effective mass.

After careful testing of the simplified procedures against the conceptually
better funded methods in equilibrium we decided to use 
Algorithm \ref{resummedmethod}
throughout this paper.

Although the different algorithms were normalized to yield equal masses in
equilibrium, it is not obvious that they will agree also for a
non-equilibrium crossover connecting two points of the $(r,h)$ plane. We
shall return to the comparison of the non-equilibrium results obtained with
different methods at the end of section IV.

\section{Passing by the critical point}

We have solved Eq.(\ref{fieldeom}), the discretized version of
Eq.(\ref{fulleom}) numerically. A starting configuration was thermalized in
presence of the initial magnetic field ($h_0=-0.2$), in such a way that a
predefined value of the average kinetic energy density was reached. This
thermal state was taken for the initial configuration when solving the field
equation with an external magnetic field tuned with constant velocity
$dh/dt$. With the parameters we used the correlation length of the initial
configuration was approximately equal to one lattice unit.

In the present investigation the same range of the parameters $r, dh/dt$ was
covered as in \cite{berdnikov00}. For each value of $r, L, a\equiv
(dh/dt)^{-1}$ runs with $\sim 20$ different equilibrium configurations were
averaged.

\begin{figure}
\begin{center}
\includegraphics[width=8cm]{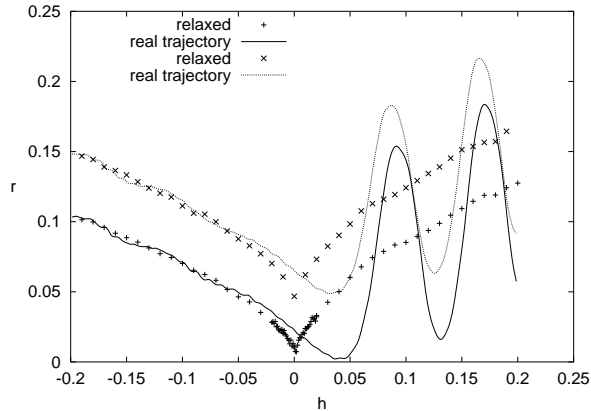}
\end{center}
\caption{
The measured trajectory in the $(r,h)$-plane for two different initial
temperatures (solid lines). The crosses show the reduced equilibrium
temperature $r(h)$ corresponding to energy density $\epsilon (h)$. The size
of the system was $L=64$, and the inverse rate of change of $h$: $a=100$.}
\label{hTroute}
\end{figure}

An approximate idea about the part of the Ising phase diagram explored in
our numerical investigation can be given by drawing the measured $(r,h)$
trajectories for different values of $r_{initial}$ and $a$ (see
Fig.\ref{hTroute}).

Due to the time dependence of the external force $h(t)$ the full energy
density $\epsilon (t)$ is not conserved, we shall parametrize it as
$\epsilon (h)$. In principle, it might depend on the parameter $a$ too, but
for its range investigated in this paper, we did not experience any
$a$-dependence of the energy density. The reduced (Ising) temperature, which 
is defined through the average kinetic energy per site,
slowly drops in the first part of the crossover. However, having passed by
the critical point ($h=0$, $r=0$) a deterministic oscillation in $r$ starts, 
which survives the averaging over the ensemble of initial thermal 
configurations. For comparison we have displayed in Fig.\ref{hTroute} also the
reduced temperatures corresponding to the equilibrium states belonging to the 
different values of the energy density $\epsilon (h)$. The latter was 
determined by stopping the variation of $h$ at a certain value $h_{stop}$, and
then relaxing the system with fixed $h=h_{stop}$ (and hence, with
conserved energy) to equilibrium. After the thermalization of this state was
complete, we have measured its equilibrium temperature. These runs were 
performed for every non-equilibrium trajectory for 40 equidistant 
intermediate values of $h_{stop}\in (-0.2, 0.2)$.

\begin{figure}
\begin{center}
\noindent\includegraphics[width=8cm]{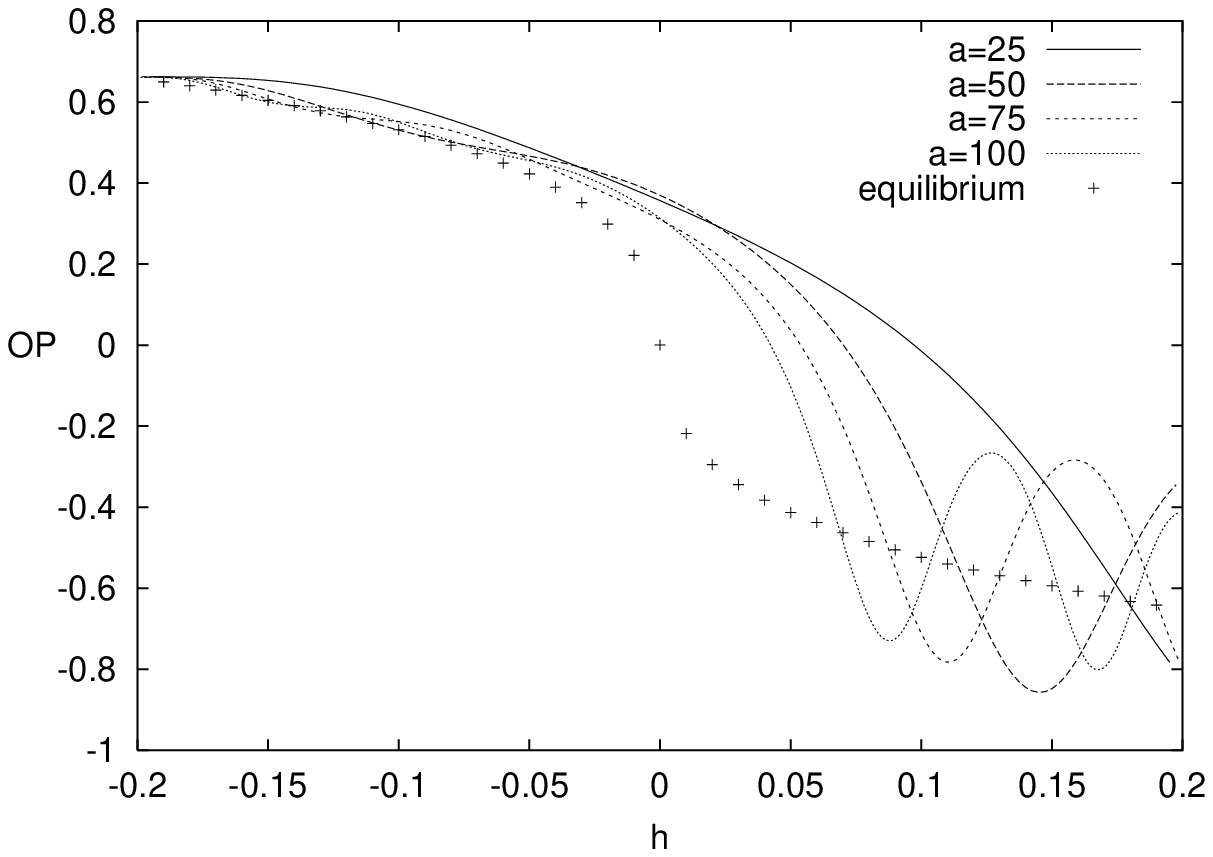}
\includegraphics[width=8cm]{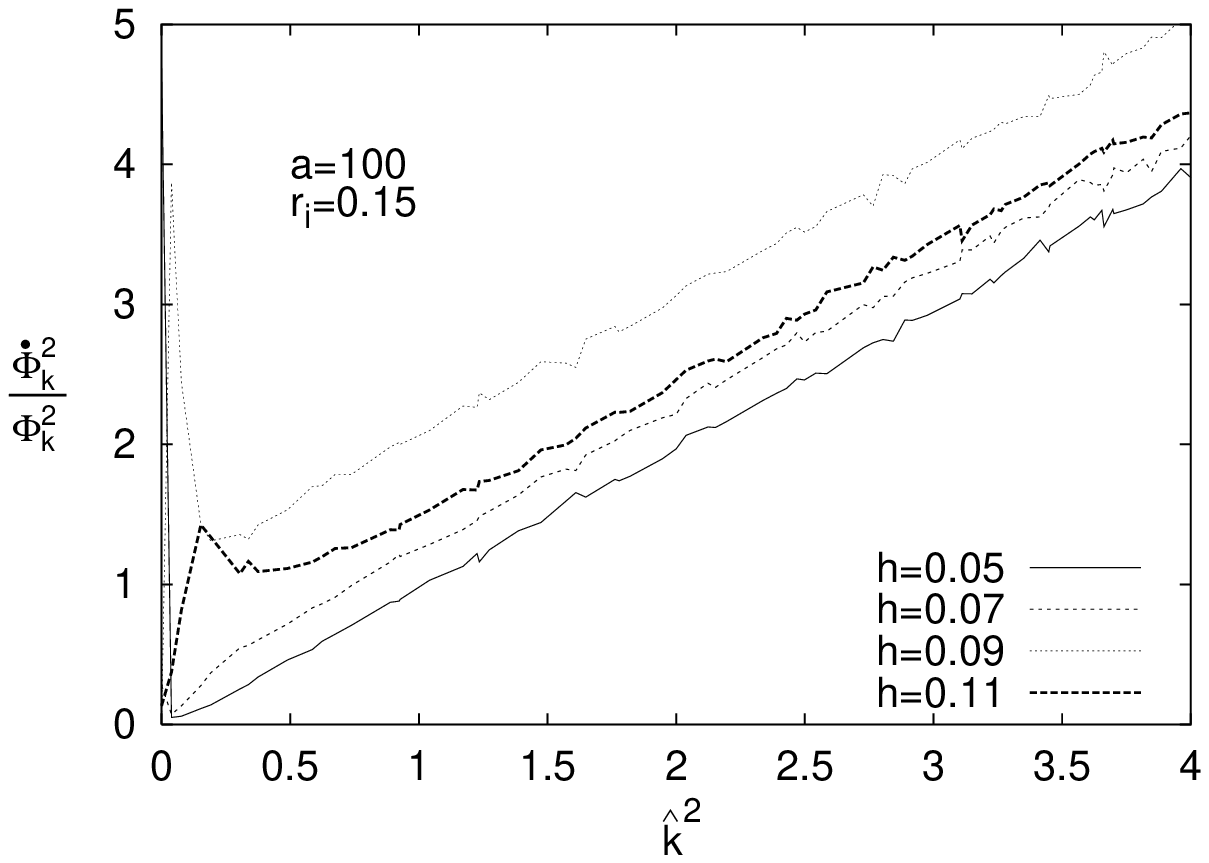}\\
\hbox to\hsize{\hfill a\pa\hfill\hfill b\pa\hfill}
\end{center}
\caption{
a\pa OP trajectories for different $dh/dt=a^{-1}$ values ($L=64$,
$r_{initial}=0.15$). The equilibrium $\bar\Phi (h)$ curve is
represented by crosses. 
b\pa Temporal variation of the left hand side of
(\ref{twopointeffmass}) synchronized by the OP oscillation.}
\label{oposc}
\end{figure}

Fig.\ref{hTroute} clearly shows how the system slows out of equilibrium. As
the critical point is approached, the thermalization time scale grows above
the time scale ($\sim ha$) of moving in the phase diagram. Thus, in the
second part of the crossover two effects seem to be present: First, the
spectral density of the configuration will correspond to the equilibrium at
an earlier $h$ value -- this means an overcooling in QCD language. Second,
the low-{\bf k} modes get highly excited and begin to oscillate with a
frequency determined by an effective mass scale.  The oscillation is
synchronously present in the entire spectrum, i.e. the kinetic energy of
each ``well-behaving'' (${\hat k}^2>1$) mode oscillates coherently. The
oscillation is not due to any direct strong coupling between the modes, but
is driven by the oscillation of the homogeneous mode ($\bar\Phi$) (see Fig
\ref{oposc}a). The UV modes adiabatically follow the slow oscillation of the
order parameter(see Fig.\ref{oposc}b), which enters parametrically through a
common effective mass in the respective equation of each mode.

One may define the error of OP trajectories
from different thermal initial conditions as the standard dispersion
of the OP($t$) values at fixed $t$.
This error comes out at $\approx0.007$ for $h<0$ and
$\approx0.02$ for $h>0$, which is about 100 times smaller than the
amplitude of the oscillations. 

For comparison the equilibrium OP values are also displayed in Fig. 
\ref{oposc}a. These values come from thermal solutions of Eq.(\ref{fulleom}) 
with the $(r,h)$ values chosen from the equilibrium points of
Fig.\ref{hTroute}.

The OP-evolution displayed in Fig.\ref{oposc}a is by itself a challenge
seeking quantitative interpretation. We shall elaborate on it in detail in
Section 4.

\section{How large the correlation length grows?}

The main physical motivation of our investigation was to answer the question
in the title of this section. We have used the fast method described in
Section 2 for deducing the actual non-equilibrium value of the correlation
length, that is the inverse $\sigma$-mass.

\begin{figure}
\begin{center}
\noindent\includegraphics[width=8cm]{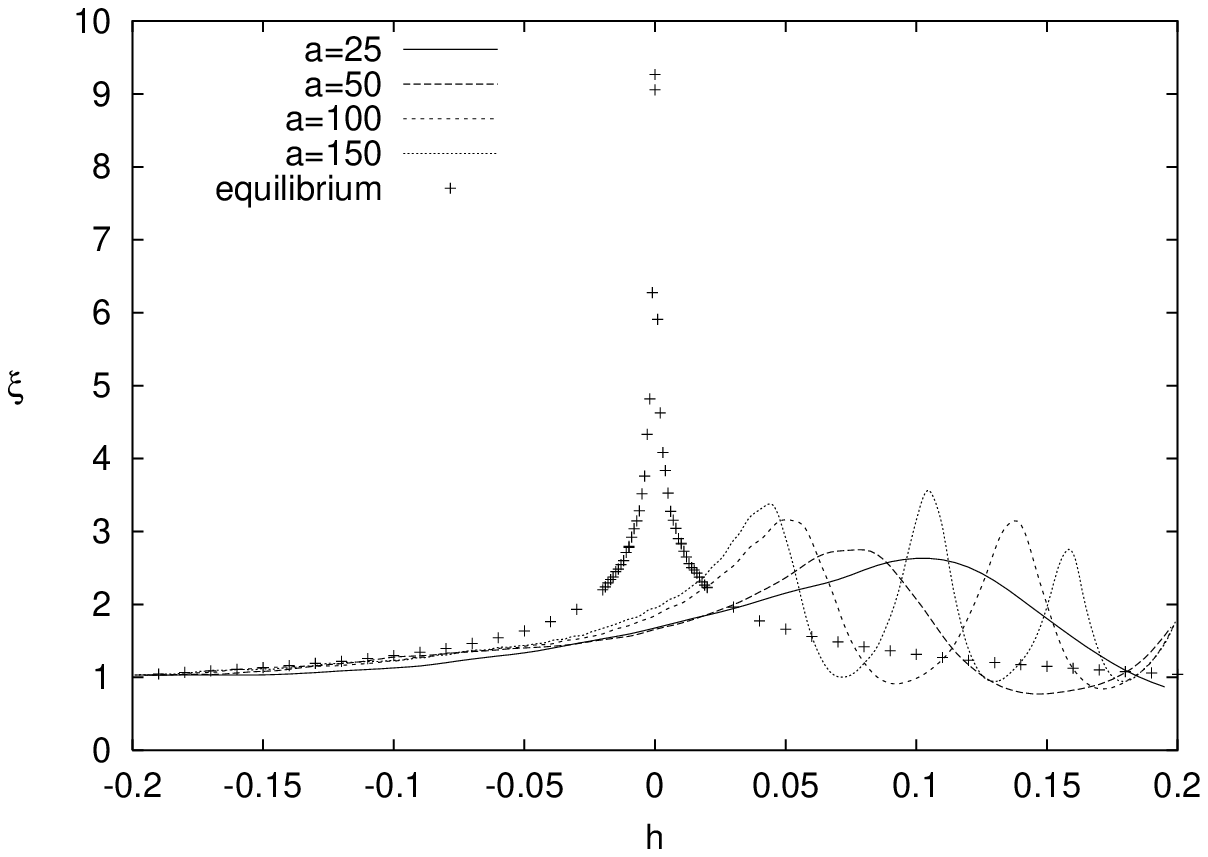}
\includegraphics[width=8cm]{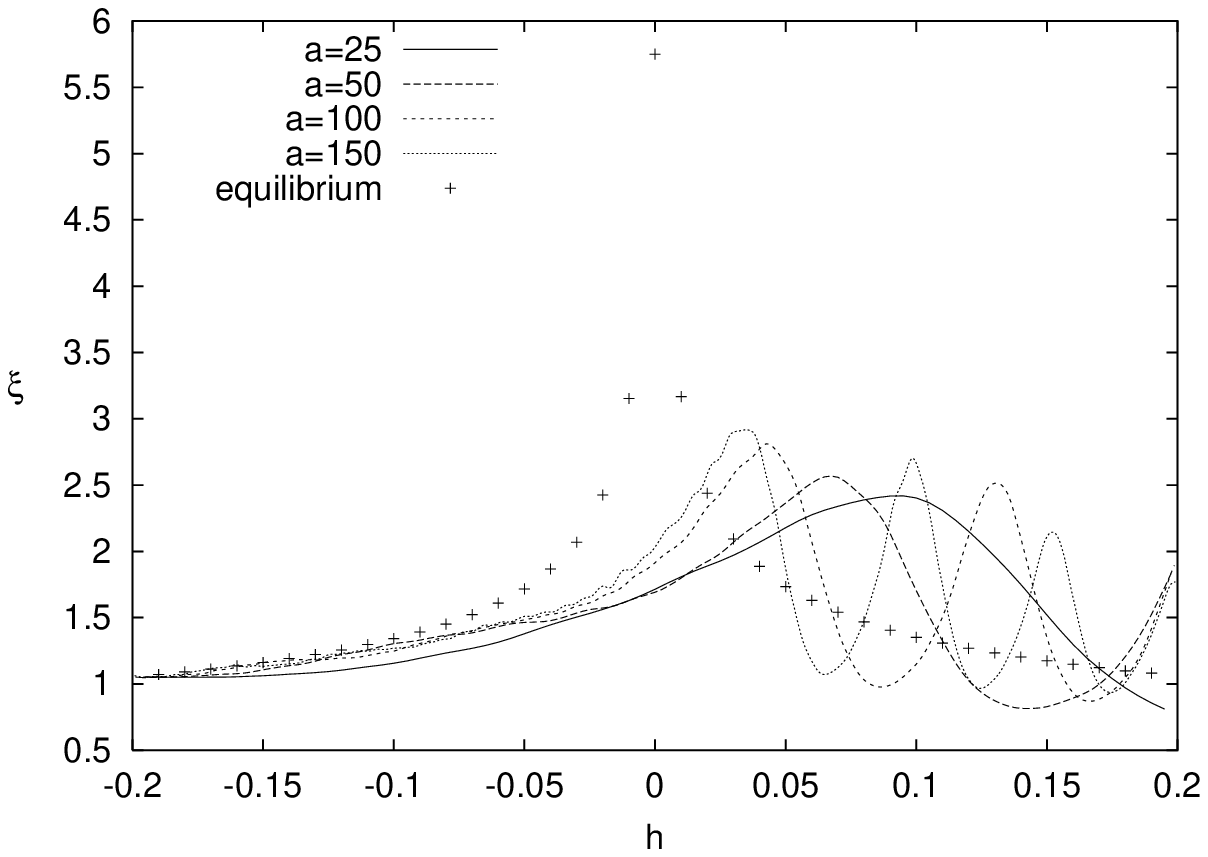}\\
\hbox to\hsize{\hfill a\pa\hfill\hfill b\pa\hfill}
\end{center}
\caption{
Evolution of correlation length during a crossover compared to the
equilibrium values along the $(r,h)$ route depicted in Fig.\ref{hTroute}.
($L=64$, $r_{i,a\pa}=0.083$, $r_{i,b\pa}=0.15$). Crosses give the results of
equilibrium measurements.}
\label{xiosc}
\end{figure}

In Fig.\ref{xiosc}a we display the $h$-history of the correlation length
obtained on a $64^3$ lattice with an initial reduced (Ising) temperature
$r_i=0.083$  for four values of the $a$ parameter. The $(r,h)$ trajectory 
starting from this $r_i$ passes the closest by the critical end point
($r_{min}=0.007$), see the lower curve in Fig.\ref{hTroute}. 
Again, the equilibrium correlation length values obtained from analyzing
long thermal time evolutions with an identical method, are displayed in the
same plot.

When compared with the estimate of \cite{berdnikov00} the most dramatic
difference is obvious: even the slowest variation of $h$ considered in BR is
in reality much too vehement. The critical slowing down of the internal
interactions pushes the system far out of the equilibrium state
corresponding to the actual magnetic field value. This is the reason that
after passing the maximum of the correlation length, reached with the
expected ``supercooling'' in $h$, $\xi$ starts to oscillate and the
correlation length very steeply drops to a minimum. This oscillation has
opposite ``phase'' if compared to the (Ising) temperature. A probable
explanation to this phenomenon is, that a shrinking correlation length means
larger $\sigma$-mass and hence an increase in the frequency of the
microscopic oscillations of the UV-modes. The energy of a weakly coupled UV
mode --- like of a tuned linear oscillator --- gets larger with the increase
of its frequency.

The qualitative picture is the same in a wider neighborhood of the critical
end-point as one can see from Fig.\ref{xiosc}b which corresponds to a
trajectory passing somewhat farther ($r_i=0.15$, the upper curve in
Fig.\ref{hTroute}).

The period of the oscillation in QCD temperature based on the $T_{QCD}-h$
correspondence proposed in \cite{berdnikov00} seems to be much smaller than
the spread of the freeze-out temperature estimates appearing in the
literature.  Therefore our main qualitative result is that the expected
increase in the coherence of the pion radiation might be missed if the
actual freeze-out would happen at a temperature slightly beyond the maximum
of the correlation length.  Ideally, accurate measurements of the freeze-out
temperatures of different meson species coupled to $\sigma$ might allow to
map out the variation of its correlation length across the hypersurfaces of
the respective ``last scatterings'' as predicted by our analysis.

One can compare the exact non-equilibrium $\xi(h,a)$ function to the
estimate of BR in some more quantitative detail. The maximal values of the
correlation length in units of the initial $\xi_0$ are in the range
$(2.5-3.5)\xi_0$. The amount of supercooling is generically larger from the
numerical solution of the $\Phi^4$ dynamics, in comparison to the result of
the first order dynamics conjectured in \cite{berdnikov00}. 

A good measure of the amount of the physical supercooling as a function of
the $h$-velocity is offered by the shift in the location of the first
maximum of the correlation length, $h(\xi_{max})$. This value turns out to
scale with the velocity of the $h$-variation:
\be
h(\xi_{max},a)=ca^{-0.5\pm 0.01}.
\ee
The values of $\xi_{max}$ themselves follow also a scaling form as
suggested in \cite{berdnikov00} on the basis of dynamical scaling
considerations. We find numerically
\be
\xi_{max}(a)=c'a^{0.211\pm 0.01}.
\ee
The value of this exponent agrees with the prediction of BR, when it
is applied to the class A. Then their prediction for the exponent yields $\nu
/\beta\delta / (1+z\nu /\beta\delta )=0.222$, with $\nu =0.630,\beta
=0.326,\delta =4.8 $ and
$z=2+(6\ln (4/3)-1)\eta ,\eta=0.0335$\cite{hohenberg77}. When the initial (Ising) 
temperature is tuned to approach the critical point the closest by 2-3\% 
from above, it turns out that the form of the $\xi (h)$ function qualitatively 
remains the same as described above. The $\xi$ values at the first maxima are
about four times larger than those taken in the first minima following them.

The second maxima of $\xi$ appearing in our numerical solution very probably
cannot be observed experimentally, since the fireball breaks up into
non-interacting mesons before reaching the corresponding low temperature.

The sensitivity of the results to the size of the system is also an
important issue, since at present the linear size of the plasma droplet is
estimated to about $6\xi_0$. Since we have chosen $\xi_0$ for the lattice  
constant, one might expect to reach the maximal allowed correlation length
in a much smaller lattice volume already. We have tested the robustness of
the above conclusions by varying the lattice size between 16 and 64. No
important variation was seen when changing the size from $L=64$ down to
$L=32$, but a 20\% drop in the maxima of $\xi$ appears when going down to
$L=16$.


The lattice we used in this investigation is rather coarse
$(|m|a_x=1)$. The idea behind this choice is that we work near the
critical point, in the scaling regime. At equilibrium when the
correlation length grows very large it is obvious that the actual
value of the lattice spacing can not matter. However, actually a
factor of 3-5 increase was experienced ``only'', therefore we have
to check that the {\it dynamical} scaling hypothesis 
\be
\xi (r,h,t)=\lambda^\nu \xi (\lambda
r,\lambda^{\nu/\mu}h,\lambda^{-\nu z}t),\qquad
\bar\Phi (r,h,t)=\lambda^{-\beta}\bar\Phi (\lambda
r,\lambda^{\beta\delta}h,\lambda^{-\nu z}t).
\label{scalrel}
\ee
is fulfilled by the solution of Eq.(\ref{fulleom}).  
The observed dynamical scaling behavior of $\xi_{max}$ and
$h(\xi_{max})$ already suggests that our system evolves in the scaling
regime. As a direct proof for this, 
we have rescaled the reduced temperature, the magnetic field and the
$a=dt/dh$ parameter in such a way that we could expect a factor of 2
increase in the correlation length if the first equation of
Eq.(\ref{scalrel}) is obeyed. The field $\Phi$ has been rescaled as
dictated by the second equation. From the rescaled equations the
evolution of the order parameter and of the inverse correlation length
has been extracted. Before the oscillation sets in the agreement with
the expectations based on the scaling hypothesis is very
convincing. It improves as the system approaches to the
critical point. This is a clean argument for the independence of our
results of the lattice spacing. This conclusion is certainly true
until the first maximum of the correlation length is reached. For the
oscillatory motion a second order dynamics is relevant, 
therefore the scaling behavior based on $z\approx 2$ should be violated.

\begin{figure}
\begin{center}
\includegraphics[width=8cm]{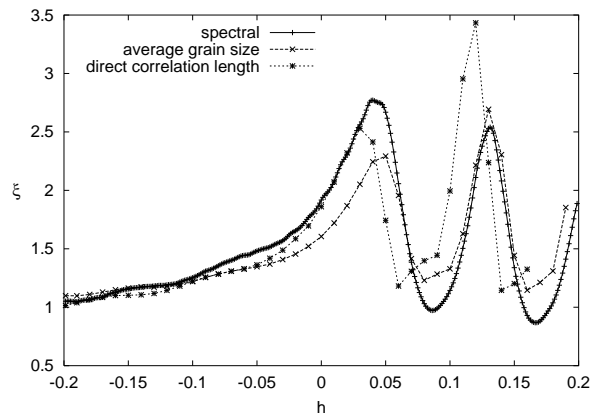}\\
\end{center}
\caption{
Comparison of the time dependence of the correlation lengths determined by
three different methods: the ``direct'' correlation length, the average
grain size and the inverse mass determined with spectral algorithms. Shown
is the evolution during the non-thermal crossover passing close to the Ising
critical end-point. ($L=32, r_i=0.15, a=100$, for definitions see the text.)}
\label{manymasses}
\end{figure}

The final question to be discussed in this section is to what extent the
observed features of the time dependence of $\xi$ depend on the algorithm of
its determination. In Fig.\ref{manymasses} a typical time evolution is
displayed for the three characteristic lengths introduced in Section II.
Using the $m_{eff}=g_i(\xi_i)$ relations determined in equilibrium
we see very good
agreement of all three before the system is slowed out of equilibrium. Next
all three enter an oscillatory regime, with about the same amplitude, but
with a certain ``phase shift''.

This observation makes it more difficult to relate the location of the first
maxima in $h$ to the freeze-out temperature. One has to work with that
length which is coupled to the microscopical mechanism producing a certain
observable. For example, the high frequency part of the $\sigma$-field is
probably well-represented by a gas of particles with the effective frequency
determined by the ``susceptibility dispersion'' method
\ref{resummedmethod}. The decay products of these hard particles will
reflect this frequency. The softest part of the pion spectra probably comes
from the coherent decay of the longest wavelength components of the field
configuration, therefore a characteristic coherence length closer to the
``grain size'' (method \ref{grainsize}) will be followed.

\section{The effective order parameter dynamics}

In statistical physics noisy first order effective equations are used for
the longest wavelength hydrodynamical modes to describe relaxation
phenomena. These equations have to be written down for the order parameter
fields and also for composite objects which correspond to densities of
conserved quantities \cite{hohenberg77}. From this viewpoint it is not clear
what could be the foundation for the proposal of BR to write down directly an
equation for the relaxation of the inverse correlation length.

We follow the more conventional path and discuss first the effective
dynamics of the order parameter. After presenting considerable evidence for
the validity of our approach to OP, we shall be able to build upon it also a
quite natural interpretation of the observed behavior of the correlation
length.
 
It is obvious that one cannot account for the oscillatory behavior of the
order parameter experienced after passing near the critical end-point just
using a first order effective dynamics. This means, that our system actually
leaves the hydrodynamical regime. Therefore, we propose to complete the
effective OP-equation with an ``acceleration'' term. For the effective free
energy we use the simplest quadratic form which corresponds to an oscillator
potential centered at the equilibrium value of the order parameter
\cite{borsanyi00}. We emphasize that no noise is introduced.

The proposed linear equation is of the form
\be
a^{-2}\ddot{\bar\Phi} (h)+a^{-1}\Gamma (h')\dot{\bar\Phi} (h)+m_{eq}^2(h')
\left(\bar\Phi (h)-\bar\Phi_{eq}(h')\right)=0.
\label{effeom}
\ee
In this fully deterministic equation a ``dot'' means derivation with respect
to $h$. In view of the relation $h=at$, the derivatives originally refer to
the time. The determination of the $h$-dependent coefficients
$m_{eq}=\xi_{eq}^{-1},\bar\Phi_{eq}$ follows the methods described in
previous sections. However, their values are taken not at the actual $h$,
but at a somewhat smaller value $h'$, which corresponds to an earlier
equilibrium state. This is the simplest way to incorporate the ``slowing out
of equilibrium'' phenomenon into the proposed equation.

The shift $h-h'$ is established by optimizing the agreement with the
measured order parameter trajectory. Three physical pictures for this shift
were tested and benchmarked by the MS deviation of the reconstructed
OP trajectory from the measured one ($\delta^2$).
In the first one a global delay parameter is introduced which
acts with equal strength before and after reaching the neighborhood of the
critical point ($\delta^2=0.0034$). In the second version one switches
on the $h$-delay only when $h\geq 0$ ($\delta^2=0.094$).
In the third version the delay grows linearly from 0 to a final value
which was fitted to the  data. This latter method produced the 
smallest MS deviation ($\delta^2=0.0029$), therefore below we shall
present results obtained with this method. The representative
$\delta^2$ values refer to a $32^3$ lattice with
$r_i=0.15, a=100$. As expected the fitted slope of the linear shift,
$h-h'=\textrm{const.}\times (h-h_0)$ clearly increases as $r_i$
gets smaller. The value of the constant coefficient changes monotonically
from 0.02 to 0.1 while $r_i$ varies from 0.2 to 0.1.
 
The reconstruction of the measured OP-trajectory turned out to be rather
insensitive to the value of the relaxation rate $\Gamma(h')$. Furthermore,
no $h'$ dependence could be observed. Chosen in the wide range
$\Gamma=0.01\dots0.6$, an acceptable agreement of the real and the
reconstructed OP-trajectory was found.

A nonperturbative, near equilibrium determination of $\Gamma$ was attempted
by stopping $h$ at a certain value $h_{stop}$ and fitting the
relaxation of $\bar\Phi$
towards equilibrium to an exponential rate. The estimate for $\Gamma$ was
found to be in the range $0.005\dots0.025$, independently of $r_i$ and $a$. 
In a model investigated previously we found that during the equilibration the
relaxation rate approaches its perturbative value strictly from below
\cite{borsanyi01}.

Eq.(\ref{effeom}) was solved with the initial values for the OP deduced
from the full system: $\bar\Phi (h=-0.2),\dot{\bar\Phi}
(h=-0.2)$. OP trajectories resulting from different realizations of
the initial thermal ensemble give slightly different initial
conditions for Eq. (\ref{effeom}). This uncertainty sets the error of
the reconstructed trajectory. (The average variance of the solution of
Eq. (\ref{effeom}) is $\sigma^2_{th}\approx 0.0001$ for $h<0$
and $\sigma^2_{th}\approx 0.0006$ for $h>0$).

\begin{figure}
\begin{center}
\noindent\includegraphics[width=8cm]{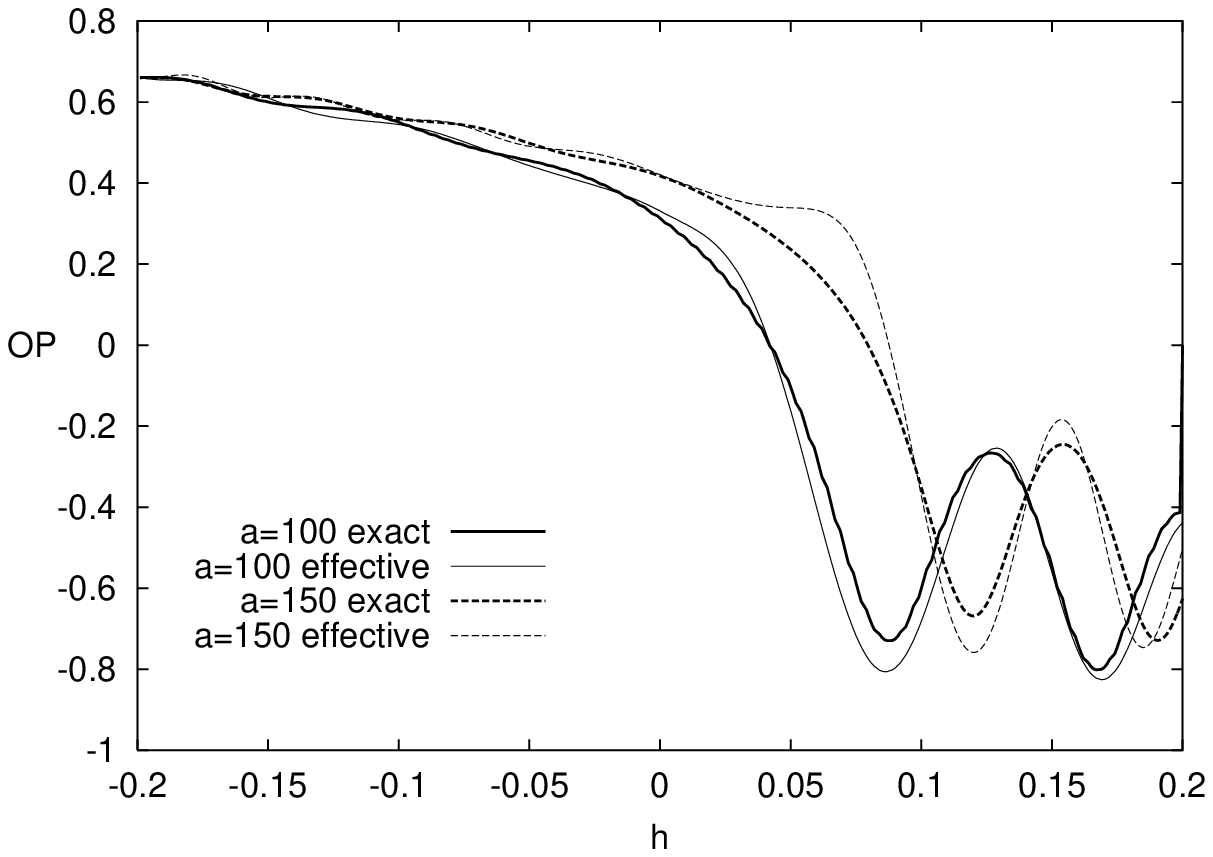}
\includegraphics[width=8cm]{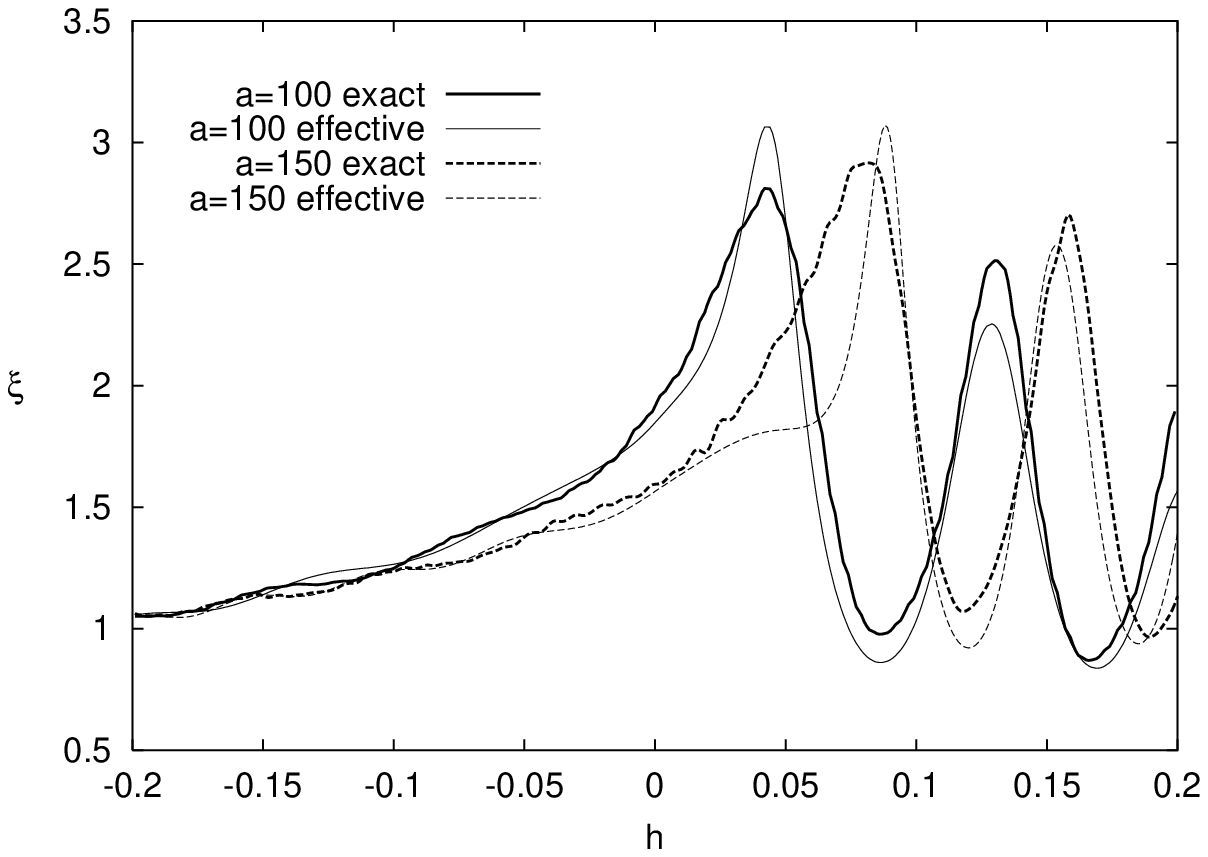}\\
\hbox to \hsize{\hfill a\pa\hfill\hfill b\pa\hfill}
\end{center}
\caption{a\pa Exact evolution of OP and the solution of
Eq.(\ref{effeom}) for two values of $a$. 
b\pa Exact evolution of the correlation length and
its values estimated with help of the effective OP-trajectory (see text) for 
two values of $a$. ($L=64$, $r_i=0.15$)}
\label{effmo}
\end{figure}

In Fig. \ref{effmo}a we compare the true and the reconstructed
OP-trajectories. The quality of the agreement somewhat fluctuates, but its 
$\delta^2$ value is less than $0.02$ in the whole $L,T,a\ge50$-region,
considered in this paper. In principle one could attempt to improve further
the analytic interpretation of the OP-dynamics by introducing a memory
kernel into the equation of $\bar\Phi$, but we believe that our proposed
equation captures the essence of the actual OP-dynamics.

With this achievement, we can return to the discussion of the non-trivial
variation of the correlation length. We build our description on our
understanding of the order parameter dynamics. 

The evolution of the system is investigated in the close neighborhood of
the critical point, in the so called scaling regime. 
In equilibrium by choosing the scale factor $\lambda =1/r$ in
Eq. (\ref{scalrel}) one finds the usual form of the  
scaling behavior of the correlation length and of the order parameter. In
Fig. \ref{hTroute} we have shown the trajectory $r=r(h)$ of the
evolution of the system for $t=ah$. If this trajectory is
simple enough, like in the left part of Fig.1, a unique $r(\xi )$
function can be extracted, for instance, from the first relation and
after its substitution into the second a functional relation 
$\xi(\bar\Phi)$ is obtained. One may then 
deduce the piecewise unique function $d\bar\Phi(\xi)/d\xi$, which can
be used to rewrite the effective equation of $\bar\Phi$ (\ref{effeom})
into an equation for $\xi(h)$. 

Motivated by this argument, we have plotted $\xi(t)$ against $\bar\Phi(t)$, 
both measured during the actual real time evolution of the system. It turns 
out that a unique functional relation $\xi(\bar\Phi)$ can be recognized 
in the regime of monotonic OP-evolution (cf. Fig. 2a). 
In the oscillatory OP-regime first one has to average the
$\xi=m_{eff}^{-1}$ values taken at different passes through
$\bar\Phi$. The function $\xi(\bar\Phi)$ is extracted only after this
step.  A very good fit valid for the whole evolution period of the form
\be
\xi^{-2}=c\bar\Phi^q+p, \qquad q\approx 2.3, \qquad c\approx 2
\ee
was obtained, with  the parameter $p$ slightly depending on the
velocity of the $h$-variation. The agreement of the measured $\xi (h)$
function with the one reconstructed by mapping the computed time
evolution of the OP using the above relation is quite spectacular 
(see~Fig.\ref{effmo}b). 

We conclude this section by discussing a more ``theoretical'' approach to
the determination of $\Gamma$, $m_{eq}$ and of $\Phi_{eq}$, by observing that
these quantities refer to (near) equilibrium situations. In our very simple
model their nonperturbative values were easily determined from the
microscopical data. In case of more realistic models, however, one may
attempt to use perturbative estimates for the masses, the damping rates and
the equilibrium order parameter. We have tested the quality of the
reconstructed OP-trajectory in the present model also when these
coefficients were taken from perturbation theory.

As it is well known even a resummed perturbation theory fails in the
vicinity of the critical temperature due to its bad behavior in the IR
regime. The divergence of the correlation length and as a consequence also
of the effective expansion parameter $\lambda T \xi$ (see for example
\cite{pietrone98}) excludes its use in the scaling regime. In a finite
system, however, with an IR cut-off $L$ we could attempt to extract the
equilibrium mass, the equilibrium OP value and the damping rate of the OP
using two-loop lattice perturbation theory. The mass comes from the second
derivative of the effective potential at the minimum, the OP is the location
of this minimum, and for the damping rate we use the formula derived in
\cite{plasmon}. The two-loop perturbative effective potential was computed
with self-consistent propagators on a finite lattice and we extracted from
it the position of the minimum and the second derivative in this point along
the ($r,h$) route shown in Fig.\ref{hTroute}. The mass used in the
propagators was determined from a one-loop gap equation.

The expression of the two-loop effective potential on lattice reads as
\begin{eqnarray}
V_{2-loop}(\bar\Phi )=V_{tree}(\bar\Phi )&+&
{T\over L^3}\sum_{\bf k}\log {\omega (\hat{\bf k},\mu )\over T}-{\lambda T^2\over
8L^6}\left [\sum_{\bf k}{1\over \omega^2(\hat{\bf k},\mu )}\right ]^2\nonumber\\
&-&{\lambda\bar\Phi^2T^2\over 12L^6}\sum_{\bf k}\sum_{\bf p}\sum_{\bf q}
{\delta^{(3)}(\hat{\bf k}+\hat{\bf p}+\hat{\bf q})\over\omega^2(\hat{\bf k},\mu)
\omega^2(\hat{\bf p},\mu )\omega^2(\hat{\bf q},\mu )},
\end{eqnarray}
where $\omega^2(\hat{\bf k},\mu )=\hat{\bf k}^2+\mu^2$. The mass was
determined from the one-loop gap equation
\be
\mu^2=m^2+{\lambda\over 2}\bar\Phi^2+{\lambda T\over 2L^3}\sum_{\bf k}
{1\over \omega^2(\hat{\bf k},\mu )}.
\ee
In view of Eq. (\ref{rescale}) we put $\lambda =6$.
 
The inaccuracy of the two-loop perturbation theory comes overwhelmingly from
the fact that for the lattice spacing considered in this paper, the estimate
of the critical temperature exceeds its value determined by us
numerically at $|m|a_x=1$ by $25\%$. Repeating both the numerical and
the perturbative calculation with $|m|a_x=0.5$ the deviation diminishes
to $5\%$. Despite all inaccuracies of the perturbation theory the
solution of Eq.(\ref{effeom}) based on the perturbative potential
$V_{2-loop}$ turns out to follow quite closely the real trajectory, though
because of the ill-determined OP values the deviation has also a systematic
error ($\delta^2\approx0.014$).

For an alternative estimate of $\xi_{eq}(h)$ we used Widom's scaling form
\cite{brezin76} like it was done by BR. This yields $\xi_{eq}(h)$ values
close to the measured equilibrium correlation length. The solution of
Eq.(\ref{effeom}) fitted to the measured curve with this
coefficient is only of slightly lower quality than the fully
nonperturbatively reconstructed one ($\delta^2\approx 0.0061$).

\section{Conclusions}

In this paper we have presented a detailed discussion of the real time
non-equilibrium evolution of the classical $\Phi^4$ field theory when it
passes nearby the critical end point in its $(r,h)$ phase diagram under the
influence of a time dependent external magnetic field. The numerical
investigation was focused on the variation of the correlation length. A
quick but accurate method for determination of the instant non-equilibrium
mass of the $\Phi$-field was employed, relying on the parametrical relation
of the spatial domain of coherence of field fluctuations to the mass.

For a wide range of the rate of variation of the magnetic field we have
experienced a slowing out of the system from the hydrodynamical regime. This
phenomenon demonstrated itself in the behavior of all quantities used for
the characterization of the system: the average kinetic energy per site, the
order parameter and the correlation length all showed oscillations when the
value of the external magnetic field moved beyond the point of
``supercooling''.

A simple effective equation was shown to describe this dramatic feature of
the OP-evolution. The coefficients of the second order differential equation
written down for the order parameter take their values from equilibrium.
They can be determined from separate simulations, but also perturbative
estimates led to reasonable description of the OP-evolution. A more
important feature of the effective equation is that the coefficient
functions should be computed for values of the external magnetic field
corresponding to some earlier stage of the evolution. The gradual increase
in this shift reflects the cumulative effect of the critical slowing down in
the proposed effective description.

The motivation for this investigation came from its possible relevance to
the $\sigma$-meson dynamics in high energy heavy ion collisions
\cite{stephanov98}. The weakest point in taking over the above features to QCD
is the lack of a quantitatively accurate mapping between QCD and
the $\Phi^4$-model in the $(T,\mu)$-plane. Still, our results can be
compared in  
a useful way with the scenario put forward in \cite{berdnikov00} for the
evolution near a hypothetical critical QCD end point.
 
In the context of heavy ion collisions, it would be interesting to see the 
effect of a coherently oscillating long wavelength $\sigma$-background with 
variable correlation length, on transversal jet quenching.

\subsection*{Acknowledgements}
The authors gratefully acknowledge important discussions with
A. Jakov\'ac and G. Veres.

\end{document}